\documentclass[journal=jctcce,manuscript=article]{achemso}
\setkeys{acs}{articletitle=true,doi=false,etalmode=truncate,maxauthors=10}
\usepackage[version=4]{mhchem} 
\usepackage{textcomp}
\usepackage{graphicx}
\usepackage{hyperref}
\usepackage{color}
\usepackage{soul}
\usepackage{dcolumn}
\usepackage{bm}
\usepackage{amsmath,amssymb}
\usepackage{natbib}
\newcommand{\doi}[1]{\href{http://dx.doi.org/#1}{\nolinkurl{#1}}}

\title{High pressure computational search of trivalent lanthanide di-nitrides}
\author{Francesca Menescardi}
\affiliation{Dipartimento di Chimica, Universit\`a degli Studi di Milano, via Golgi 19, 20133 Milan, Italy}
\affiliation{Consiglio Nazionale delle Ricerche, Istituto di Scienze e Tecnologie Chimiche (CNR-SCITEC), via Golgi 19, 20133 Milan, Italy}

\author{Emma Ehrenreich-Petersen}
\affiliation{Department of Chemistry and iNANO, Aarhus University, Langelandsgade 140, 8000 Aarhus C, Denmark}

\author{Davide Ceresoli}
\email{davide.ceresoli@cnr.it}
\affiliation{Consiglio Nazionale delle Ricerche, Istituto di Scienze e Tecnologie Chimiche (CNR-SCITEC), via Golgi 19, 20133 Milan, Italy}

\begin{document}

\begin{tocentry}
\centering
\includegraphics[height=3.5cm]{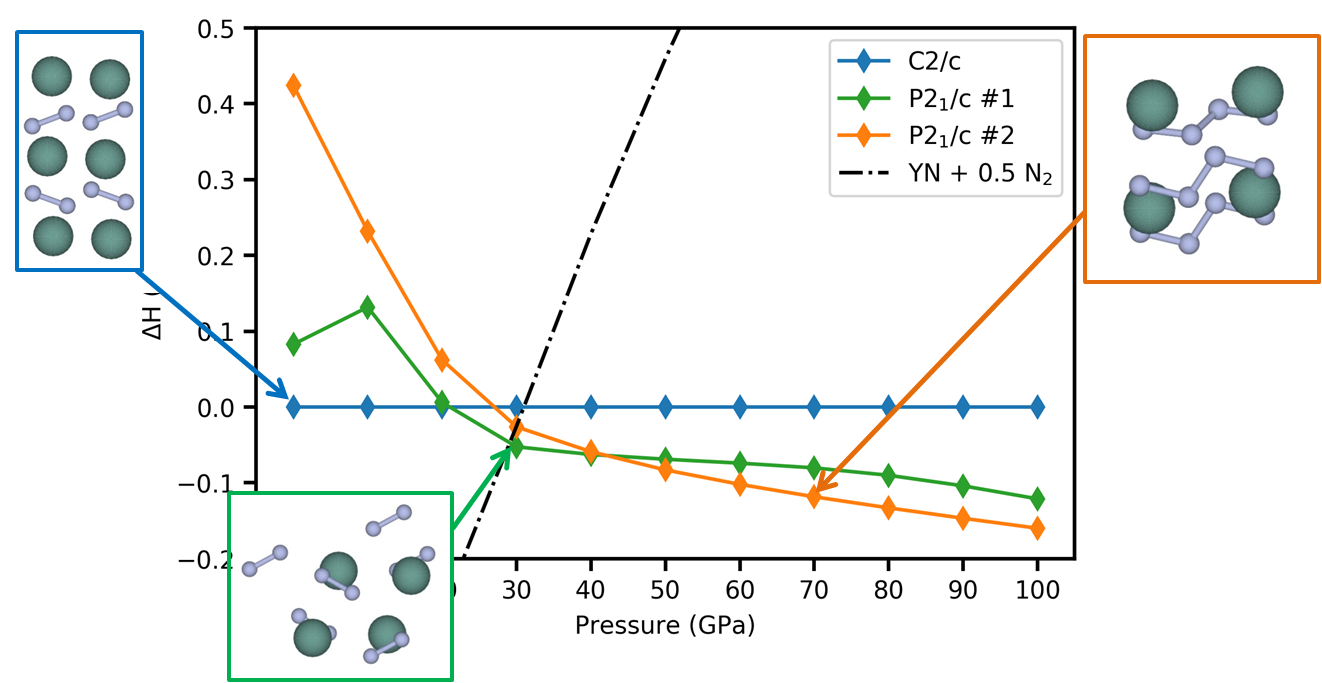}
\end{tocentry}

\begin{abstract}
Transition metal nitrides have attracted much interest of the scientific community for
their intriguing properties and technological applications. Here we focus on yttrium
dinitride (YN$_{2}$) and its formation and structural transition under pressure. We employed
a fixed composition USPEX search to find the most stable polymorphs. We choose yttrium as
a proxy for the lanthanide series because it has only $+3$ oxidation state, contrary
to most transition metals. We then computed thermodynamic and dynamical stability
of these structures compared to the decomposition reactions and we found that the compound
undergoes two structural transitions, the latter showing the formation N$_{4}$ chains.   
A closer look into the nature of the nitrogen bonding showed that in the first two structures,
where nitrogen forms dimers, the bond length is intermediate between that of a
single bond and that of a double bond, making it hard to rationalize the proper oxidation
state configuration for YN$_{2}$. In the latter structure where there is the formation of
N$_{4}$ chains, the bond lengths increase significantly, up to a value that can be justified
as a single bond. Finally, we also studied the electronic structure and the dynamical stability
of the structures we found.

\end{abstract}

\maketitle

\section{Introduction}
The intriguing properties and potential vast application of transition metal nitrides (TMNs)
have attracted the interest of the scientific community in recent
years~\cite{Steele2017,Wu2018,Xia2019,Jiao2020,Wang2020,Zhou2020a,Zhou2020b}.
In general, nitrides are much less studied than the corresponding oxygen-based compounds, but
have been shown to give rise to ultrahard~\cite{Auffermann2001,Weihrich2003,Weihrich2003a,Wang2011,Bhadram2016,Alkhaldi2019,Pillai2020}
and superconducting~\cite{Shy1973,Wang2015} materials that can find application in modern
technologies as field effect transistors, p-n junctions and energy storage devices~\cite{Zhou2020a,Zhou2020b}.

In this work we focus on binary nitrides with a formally trivalent cation, such as a rare-earth
ion. Nitrogen-rich compounds that can be synthesized under high pressure, exhibit intriguing
properties compared to the mononitrides. In fact, besides the excellent mechanical properties
that arise, when M$-$N$_x$ compounds are synthesized under pressure, they tend to form fairly
long chains of nitrogen atoms. The consequent formation of high energy, single N$-$N bonds is
an effective way to store chemical energy~\cite{Steele2017}. This is not surprising as nitrogen itself is
know to polymerize at 110~GPa and 2000~K, forming the so called \emph{cubic gauche} structure.
Very recently, a new polymeric polymorph resembling black phosphorous was synthetized.~\cite{Laniel2020}
Moreover, it is known that the alloying of nitrogen with alkali or with transition metal ions
is an effective way to decrease the polymerization pressure~\cite{Xia2019}.

The mononitrides with a trivalent rare-earth cation are reported to form simple packed structures
under pressure (B1, B2, B10)~\cite{Yang2010,Nielsen2017,Emma2020} with a suggested formal oxidation
state configuration
M$^{3+}$N$^{3-}$, resulting in small to wide band-gap semiconductors~\cite{TieYu2006}. On the contrary,
several valence configurations have been proposed~\cite{Wessel2010} to explain the nature of the
nitrogen chemical bonding and its relation to the physical properties of MN$_2$ materials.
Tetravalent cations form compounds with formal oxidation M$^{4+}($N$_{2})^{4-}$ that 
are called \emph{pernitrides} and are characterized by a N$-$N bond length of $\sim$1.42~\AA.
Divalent cations form compound with formal oxidation state M$^{2+}($N$_{2})^{2-}$
that are called \emph{dinitrides} and display a significantly shorter N$=$N bond length of
about 1.23~\AA. The case of a trivalent rare-earth
cation does not seem to fit in any of the two definitions, and the understanding of the nature
of N..N bond in this case is not so intuitive. In a recent work, Wessel et al.~\cite{Wessel2010} 
rationalized the nature of the nitrogen bonding in LaN$_{2}$ compound, suggesting a possible
mechanism for the formation of this non-trivial structure. Indeed, forcing the N$_2$ moiety
away from the $-2$ and $-4$ oxidation states, could be an effective way towards the formation
of long nitrogen chains.

To explore these concepts we studied the structural, electronic and formation properties
of yttrium dinitride (YN$_2$) under pressure. We choose yttrium as a proxy for lanthanides,
because it is always
trivalent, avoiding the multiple valence character of some lanthanides ions. Moreover, from a
computational point of view, yttrium displays empty highly-localized $4f$ orbitals, thus avoiding
complications such as magnetism and strong correlation, that arise when in presence of
partially filled $f$ orbitals.

We applied an ab-initio crystals structure prediction (CSP) method~\cite{OganovBook2018} to
search for the most stable YN$_2$ polymorphs up to 100~GPa. We computed the thermodynamic
and dynamic stability of the obtained polymorphs down to ambient pressure and with respect
to the decomposition reaction YN$_2\rightarrow$YN$+\tfrac{1}{2}$N$_2$. Finally we compared
the nitrogen bond length and bulk modulus to a set of existing compounds and in order to
rationalize the nature of chemical bonds and to understand which formal valence better
describes the chemistry of these compounds.

\section{Computational methods}
We performed a fixed composition USPEX (v9.4.4) search on four YN$_2$ unit formula. The initial
USPEX population was composed of 40 randomly generated structures and we allowed up to 25
generations, employing the standard genetic algorithms (heredity, random generation, soft mutation,
permutation and lattice mutation)~\cite{Oganov2018,Oganov2019,Oganov2019a}. The external
pressure was set to 100 GPa, about double the highest pressure that is expected to lead to
the experimental synthesis of pernitrides. To reduce the computational cost, we used the local-basis
code SIESTA~\cite{Artacho2008}, with norm-conserving pseudopotentials from the Martins-Trouiller
table and the SZP basis set with a 50~meV energy shift. The mesh cutoff was 250~Ry. We used the
PBE exchange correlation functional~\cite{Perdew1996}.

At the end of the USPEX run, we selected the 20 most stable structures, after pruning those
we found equivalent by symmetry. These structures were further relaxed down to ambient pressure
in steps of 10~GPa with the plane-wave pseudopotential code Quantum Espresso~\cite{Giannozzi2009,Giannozzi2017}.
We used the PBE functional, ultra-soft pseudopotentials from the GBRV library~\cite{Garrity2014},
wave function/density cutoffs of 45/450~Ry and up to 6$\times$6$\times$6 k-points. The
electronic density of states was computed on a finer k-point mesh. The phonon density of states
were computed using the density functional perturbation theory (DFPT) method~\cite{Baroni2001}
with a density cutoff of 900~Ry.

\section{Results}
\subsection{\protect{YN$_2$} structures}
\begin{figure}
    \centering
    \includegraphics[width=\columnwidth]{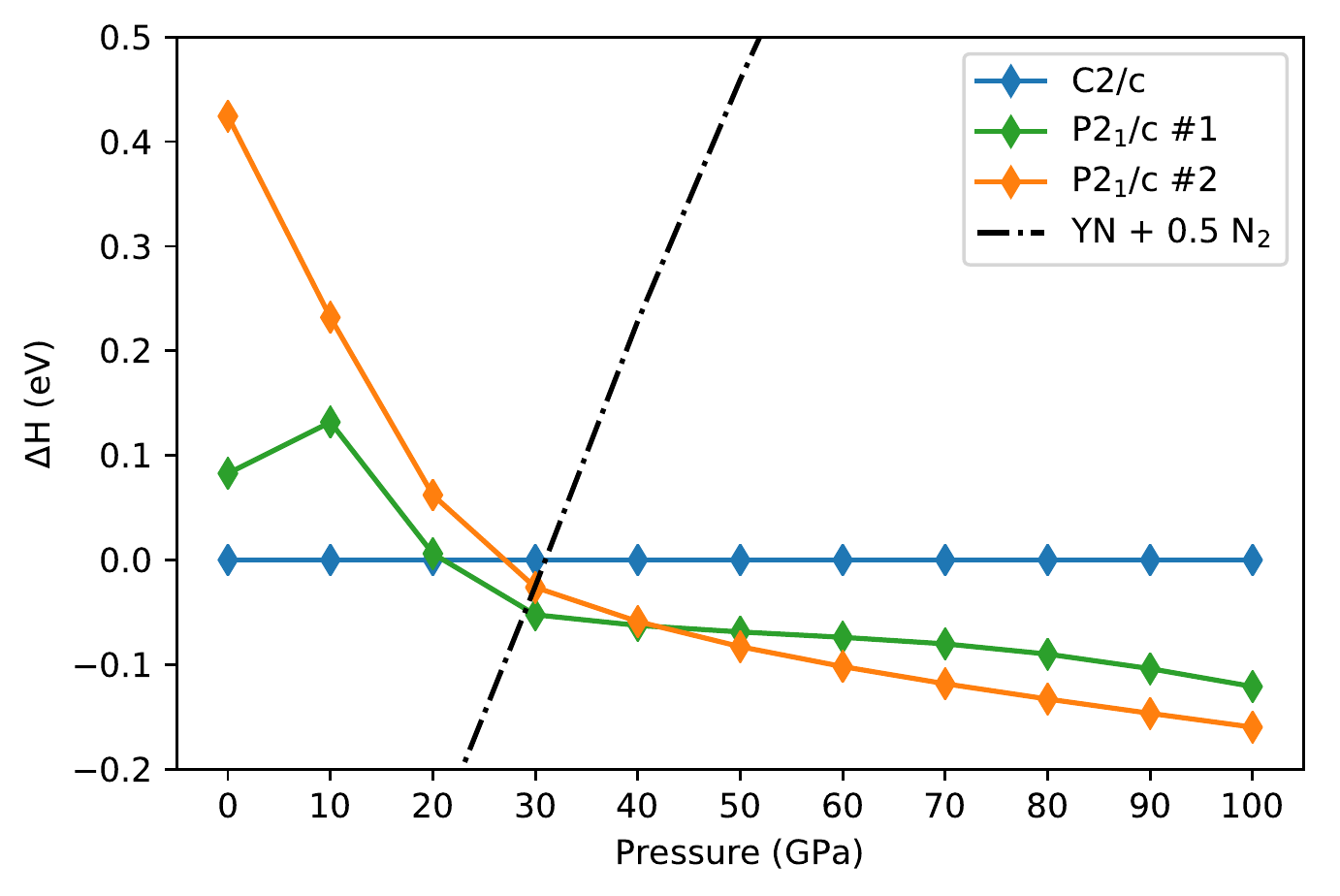}
    \caption{Calculated enthalpy as function of pressure, of the three most stable YN$_2$
    polymorphs, relative to the enthalpy of the ThC$_2$ structure. The dashed line marks
    the enthalpy of bulk YN (B1 structure) plus $\tfrac{1}{2}$N$_2$ ($\epsilon$-N$_2$ structure).}
    \label{fig:enthalpy}
\end{figure}

Within the maximum pressure range of 100~GPa, we found three stable polymorphs among all the
structures produced by USPEX (Fig.~\ref{fig:enthalpy}). The other higher energy polymorphs,
not reported in this paper, usually differ from the the most stable one, by the stacking
of the layers. We note that even though the polymorphs appear like layered structures, the
inter-layer interaction is not van der Waals or dispersion. Hence, the layer stacking has a
strong impact on the enthalpy of the system.

\begin{figure}
    \centering
    \begin{tabular}{ccc}
    \includegraphics[width=0.3\columnwidth]{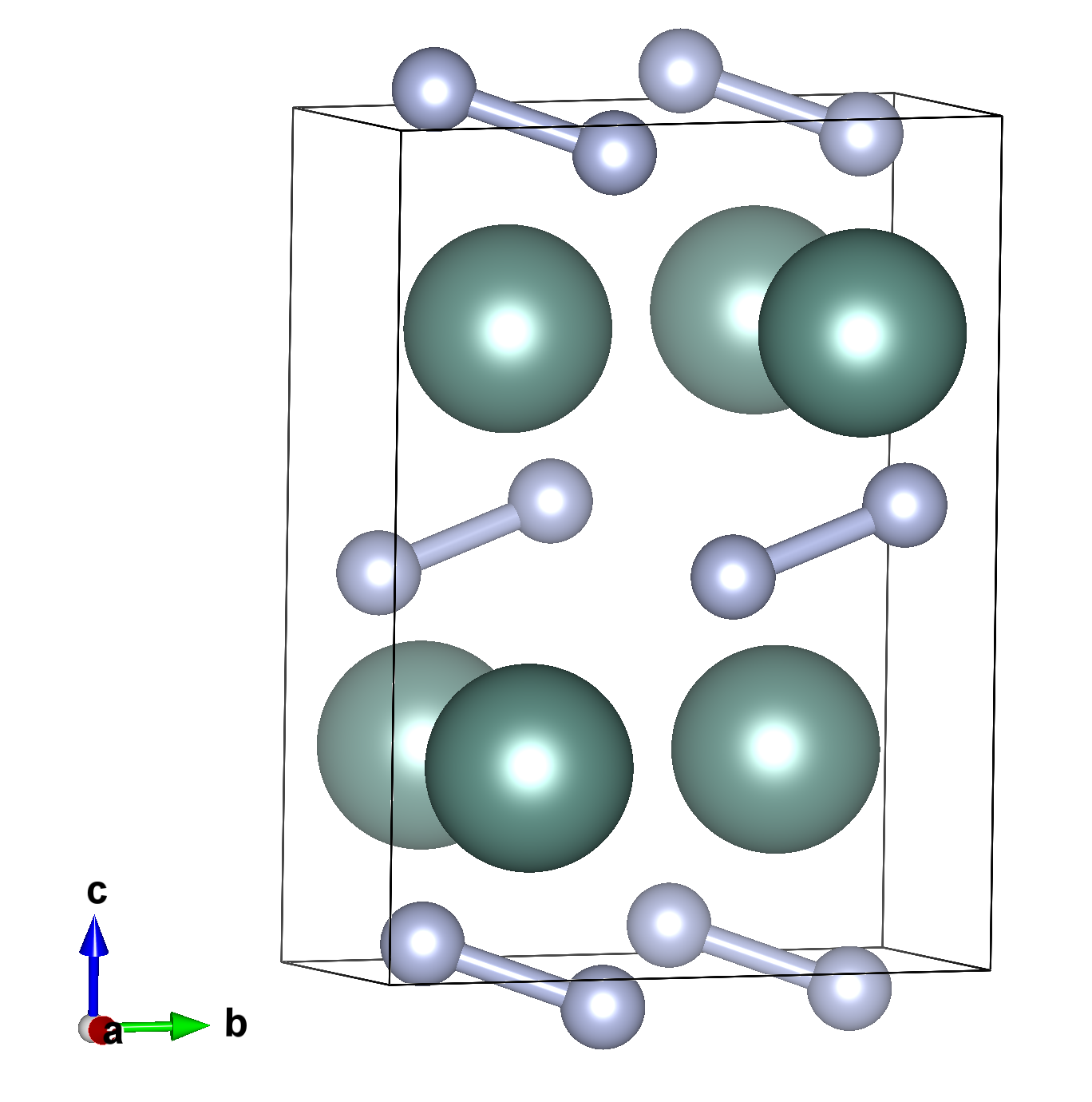} &
    \includegraphics[width=0.3\columnwidth]{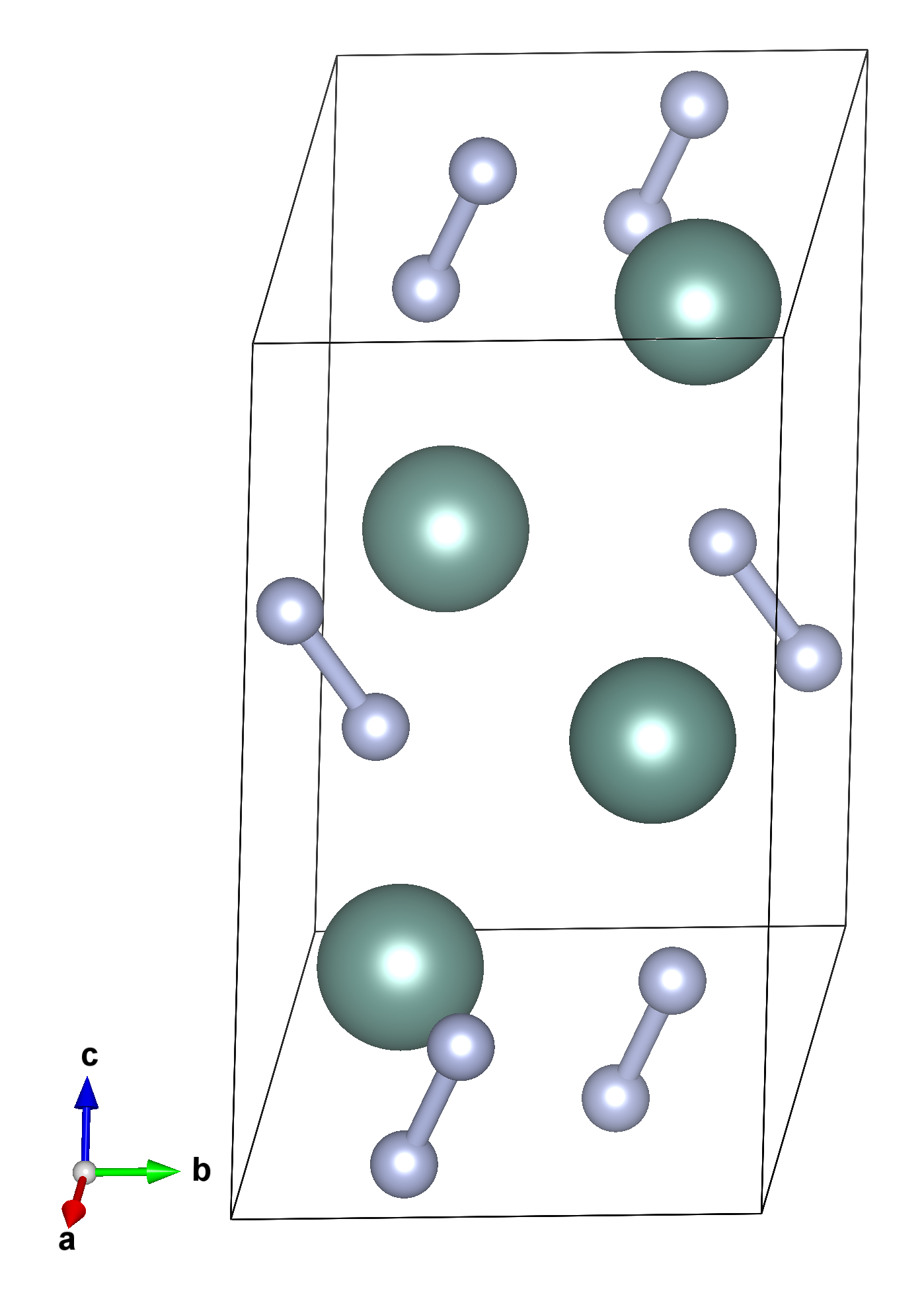} &
    \includegraphics[width=0.3\columnwidth]{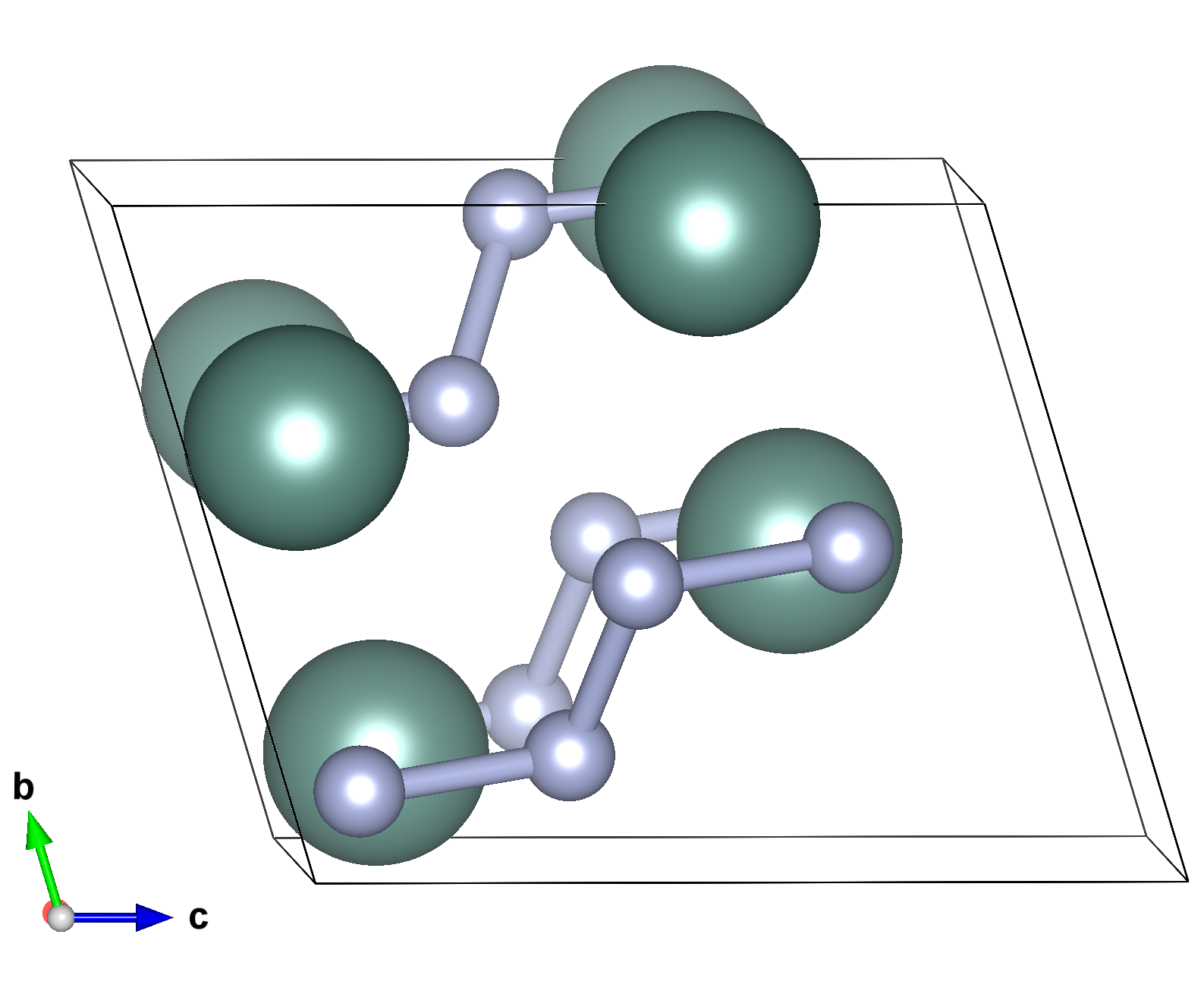} \\
    (a) & (b) & (c)    
    \end{tabular}
    \caption{Fully optimized YN$_2$ structures. (a) C2/c (ThC$_2$) structure at 0~GPa.
    (b) P2$_1$/c\#1 structure at 30~GPa. (c) P2$_1$/c\#2 structure at 50~GPa. The green
    spheres are the Y ions, the small gray spheres are the N ions.}
    \label{fig:structures}
\end{figure}

At ambient pressure, we predict that YN$_2$ will adopt the ThC$_2$ (space group C2/c~\cite{Guo2017}) crystal structure
(Fig.~\ref{fig:structures}a), characterized by alternating layers of Y and N$_2$ dimers.
Upon increasing the pressure the structure undergoes a structural phase transition at GPa
into a more compact
P2$_1$/c\#1 structure (Fig.~\ref{fig:structures}b), still with N$_2$ dimers, but with
no evident alternate layers of Y and N$_2$. Then, at 40 GPa, it turns into another
monoclinic structure with
the same symmetry (P2$_1$/c\#2), but with by N$_4$ moieties (Fig.~\ref{fig:structures}c).
The N$_4$ chains persist up to the highest investigated pressure of 100~GPa. The lattice
parameters and the Wyckoff positions of the structures at 0, 30, 50 and 100 GPa are listed
in Tab.~\ref{tab:structures}.

\begin{figure}
    \centering
    \includegraphics[width=\columnwidth]{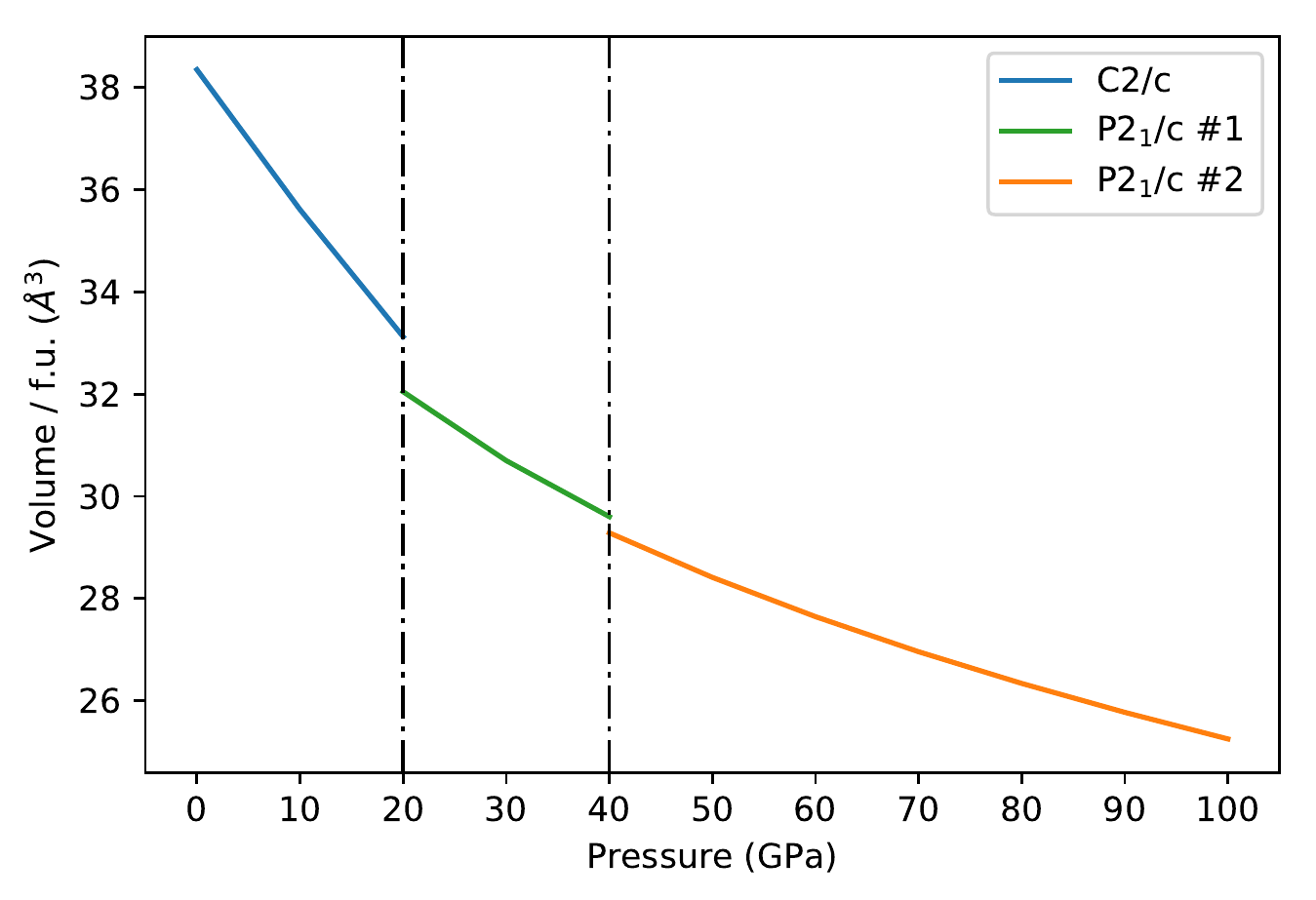}
    \caption{Calculated volume per formula unit as function of pressure. The vertical lines
    indicate the transition pressures.}
    \label{fig:volume}
\end{figure}

\begin{table}
    \centering
    \begin{tabular}{lllcc}
    \hline\hline
    Phase & Pressure & Lattice & Sites & Coordinates \\
    \hline
    C2/c   & 0 GPa         & a=6.0492~\AA                 & 4e    & Y: 0.0000, 0.2276, 0.2500 \\
    (s.g. 15) &            & b=4.2245~\AA                 & 8f    & N: 0.2260, 0.5389, 0.5389 \\
           &               & c=6.2247~\AA                 &       &  \\
           &               & $\beta=$74.64$^\circ$        &       &  \\
    \hline
    P2$_1$/c\#1 & 30 GPa   & a=4.3842~\AA                 & 4e    & Y: 0.6205, 0.7824, 0.9175 \\
    (s.g. 14) &            & b=4.7349~\AA                 & 4e    & N: 0.0289, 0.9216, 0.3154 \\
           &               & c=7.7441~\AA                 & 4e    & N: 0.7500, 0.2130, 0.4771 \\
           &               & $\beta=$134.50$^\circ$       &       & \\
    \hline
    P2$_1$/c\#2 & 50 GPa   & a=5.6303~\AA                 & 4e    & Y: 0.7723, 0.0131, 0.1563 \\
    (s.g. 14) &            & b=4.6147~\AA                 & 4e    & N: 0.8210, 0.5166, 0.1802 \\
           &               & c=4.5694~\AA                 & 4e    & N: 0.4336, 0.0820, 0.3679 \\
           &               & $\beta=$108.80$^\circ$       &       & \\
    \hline
    P2$_1$/c\#2 & 100 GPa  & a=5.4306~\AA                 & 4e    & Y: 0.7695, 0.0232, 0.1518 \\
    (s.g. 14) &            & b=4.4185~\AA                 & 4e    & N: 0.8262, 0.5283, 0.1761 \\
           &               & c=4.3994~\AA                 & 4e    & N: 0.4333, 0.0844, 0.3666 \\
           &               & $\beta=$106.90$^\circ$       &       & \\
    \hline\hline
    \end{tabular}
    \caption{Lattice parameters, space group and Wyckoff positions of the YN$_2$ structures, at
    selected pressures.}
    \label{tab:structures}
\end{table}

Fig.~\ref{fig:volume} shows the volume of the structures considered in this work. Here,
volume is seen to decrease as a consequence of the great pressure applied. In addition,
a noticeable volume drop  is seen for both the first structural transition (20 GPa, 3.25\%)
and the second structural transition (40 GPa, 1.07\%). Clearly, the volume drop of
the second transition is lower and both structural transitions are probably of first-order.  

In Fig.~\ref{fig:enthalpy} we also report the enthalpy of bulk YN (B1 structure) $+$
$\tfrac{1}{2}$N$_2$. The USPEX search on the mononitride YN found that the B1 structure
is by far the most stable polymorph in the pressure range. For sake of simplicity,
we took the $\epsilon-$N$_2$ polymorph to model solid nitrogen in the whole pressure range
as it is reported to be the most stable polymorph between 13~GPa and 69~GPa. We found
that in this pressure range, below 30~GPa the P2$_1$/c\#1 is not thermodynamically
stable and it decomposes into YN and N$_2$.

We also run a USPEX calculation with three unit formula but the structures obtained are
less thermodynamically stable than the structures with four unit formula. The reason is
that the former are characterized by a N$_2$ and a N$_4$ moiety. These ``4+2'' structures
could be considered as intermediate between the P2$_1$/c~\#2 and the P2$_1$/c~\#1, where
the N$_4$ chains break into pairs of N$_2$ molecules as the pressure is lowered.
Interestingly, we didn't find nor any N$_3$ moiety neither cyclic nitrogen species
stable enough to be competitive with the structures reported so far.

\subsection{Dynamical stability}
\begin{figure}
    \centering
    \begin{tabular}{cc}
    (a) & \\ & \includegraphics[width=0.5\columnwidth]{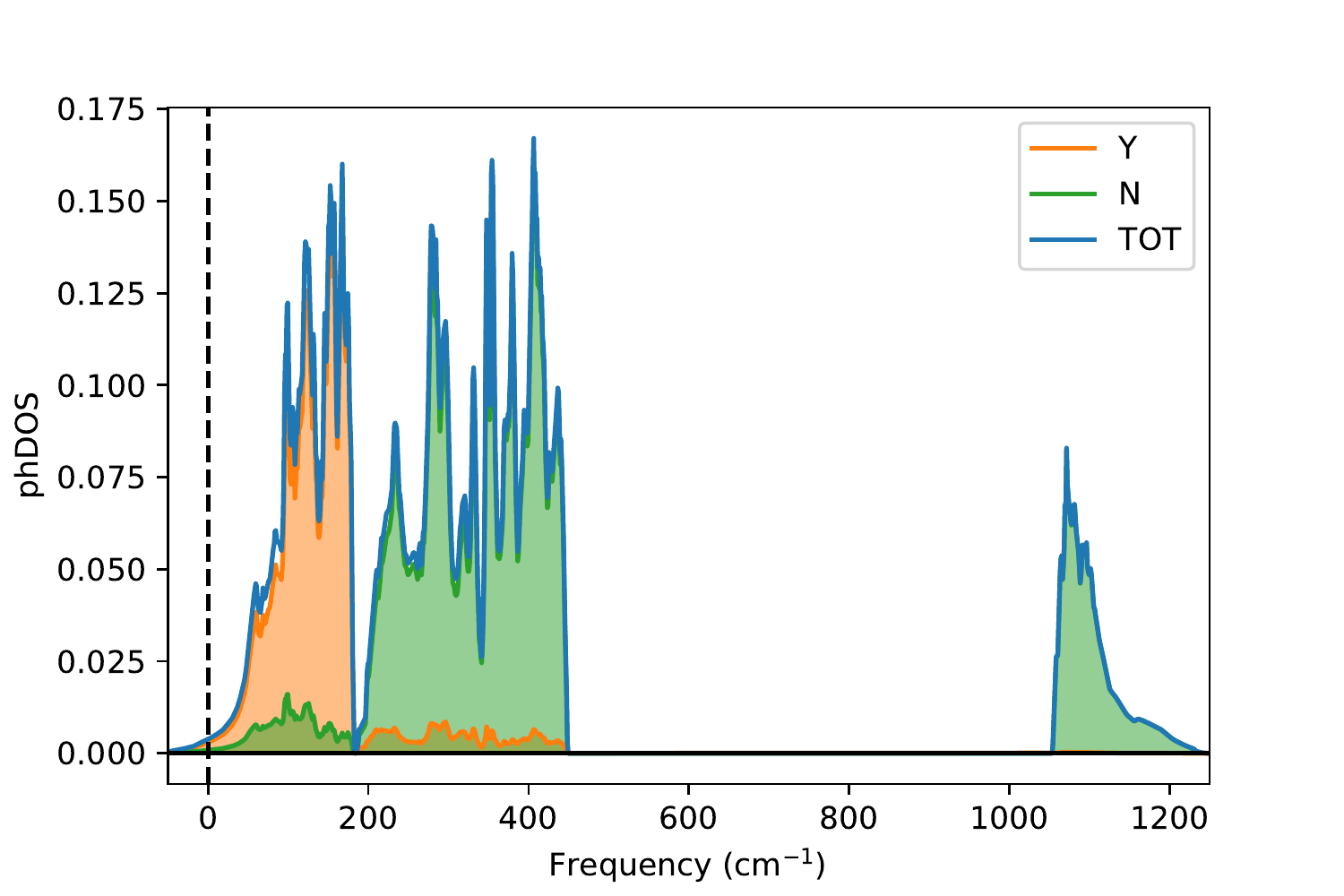} \\
    (b) & \\ & \includegraphics[width=0.5\columnwidth]{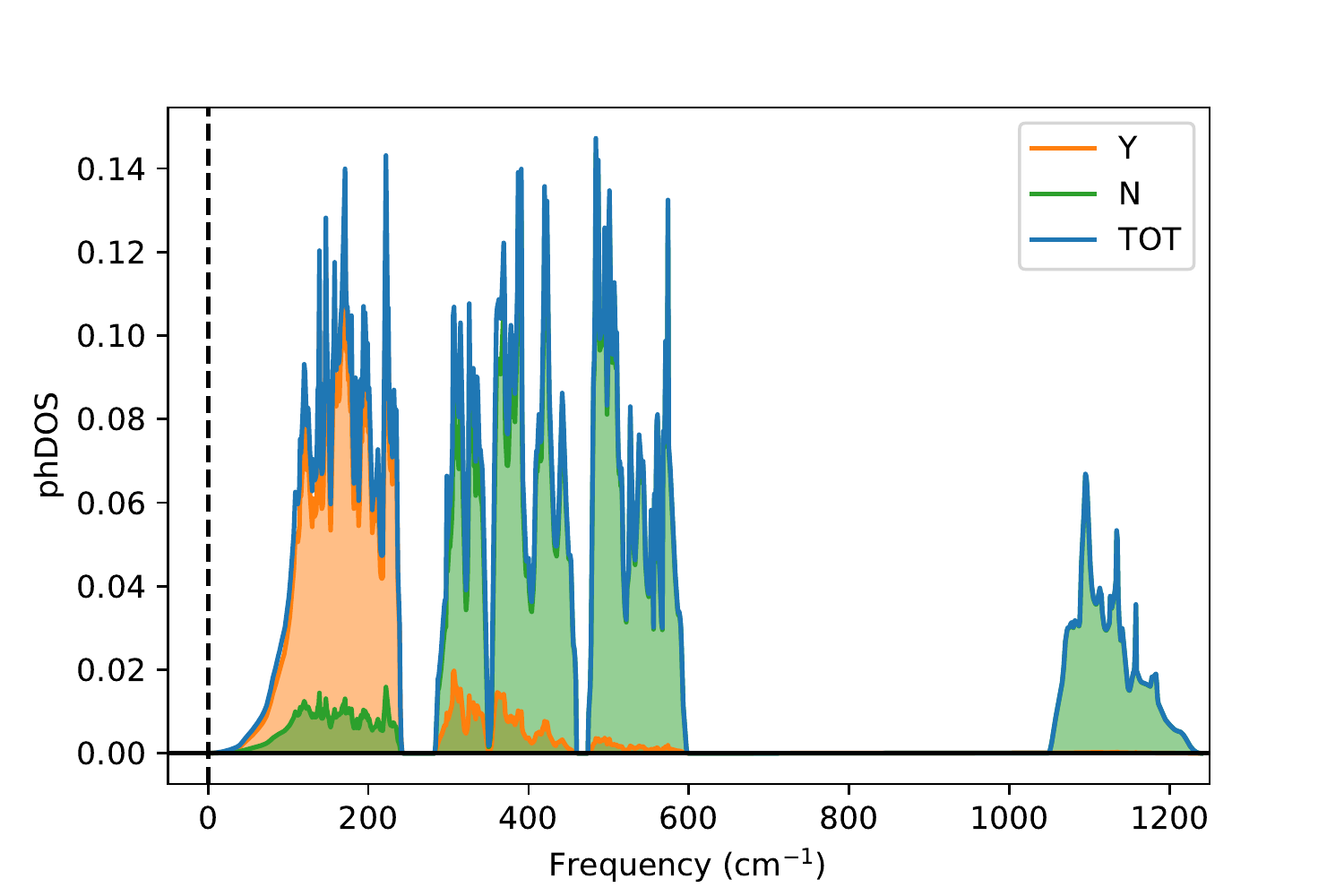} \\
    (c) & \\ & \includegraphics[width=0.5\columnwidth]{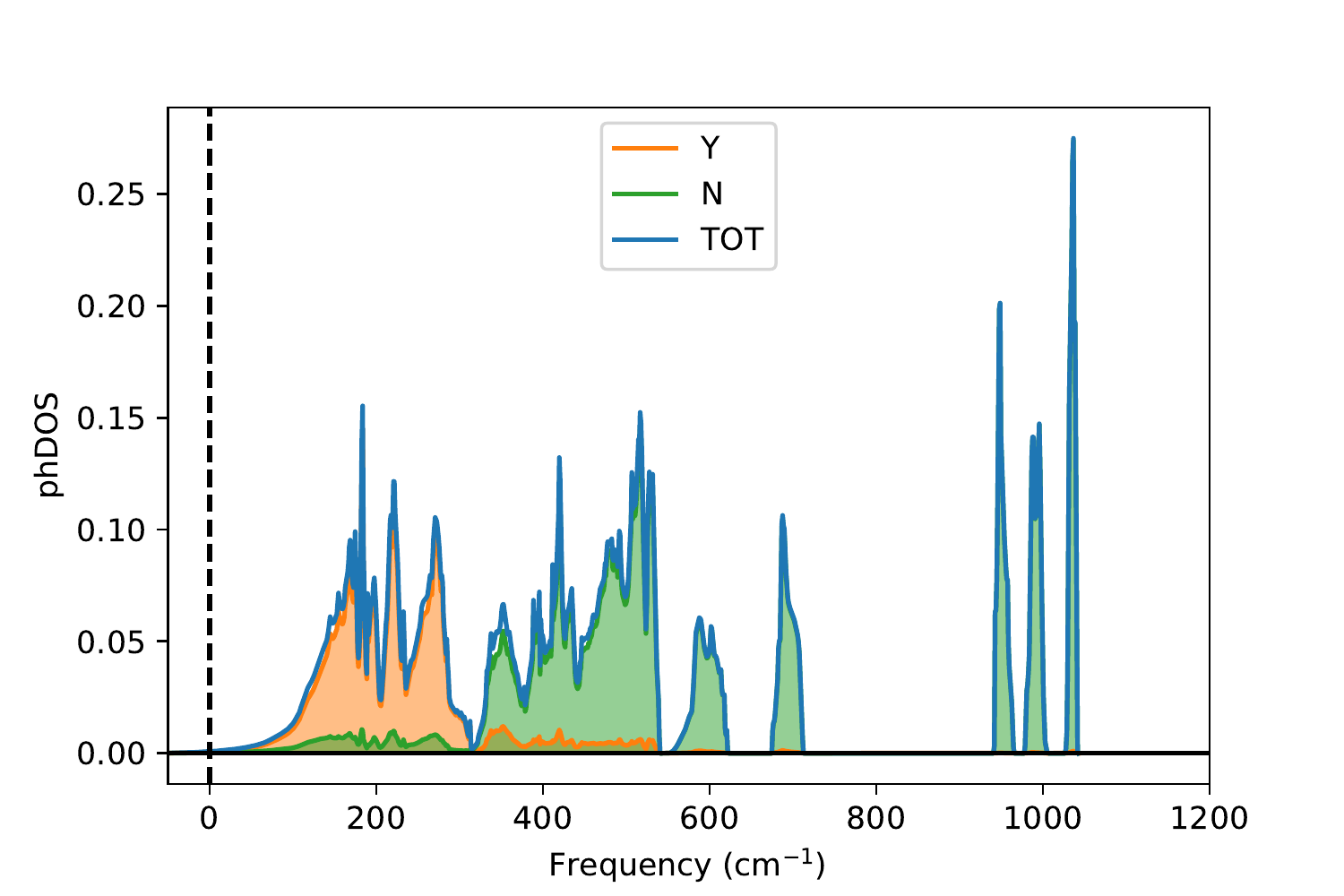} \\
    \end{tabular}
    \caption{Calculated phonon density of states of the thermodynamically stable structures.
    (a) C2/c phase at 0~GPa; (b) P2$_1$/c\#1 phase at 30 GPa; (c) P2$_1$/c\#2 phase at 50~GPa.
    The Y and N contributions to the phonon density of states are indicated by the orange
    and green shaded areas, respectively.}
    \label{fig:fononi}
\end{figure}

To establish that the predicted structures are dynamically stable, we calculated the
phonon dispersion on a $q$-point mesh of 3$\times$3$\times$3. In Fig.~\ref{fig:fononi}
we report just the phonon density of states, rather than the phonon dispersion
since all structures are monoclinic. From Fig.~\ref{fig:fononi} we found that the
ThC$_2$-structure is
dynamically unstable at ambient pressure. Inspection of the unstable phonon modes,
reveals that the stable structure would display a long wavelength modulation
of the relative distance between the Y and N$_2$ sublattices. However, the P2$_1$/c\#1
structure is thermodynamically unstable against the decomposition, as shown previously.
Therefore, this lattice modulation, akin to
a charge density wave instability would be extremely difficult to observe in experiments.
On the contrary, the other two structures are dynamically stable since all the phonon
frequencies are real and positive. An interesting feature is that the nitrogen contributes
mainly to the high energy optical modes at frequencies 1000~cm$^{-1}$ and above.
In the low energy region, there is still a noticeable separation between Y and N vibrations.
Indeed the heavy Y ion contributes mainly to the lowest energy phonon (below 200~cm$^{-1}$),
whereas the intermediate energy region is largely characterized by vibrational modes involving
nitrogen only.

\subsection{Electronic structure and chemical bonding}
\begin{figure}
    \centering
    \begin{tabular}{cccc}
    (a) & (b) & (c) & (d) \\
    \includegraphics[width=0.2\columnwidth]{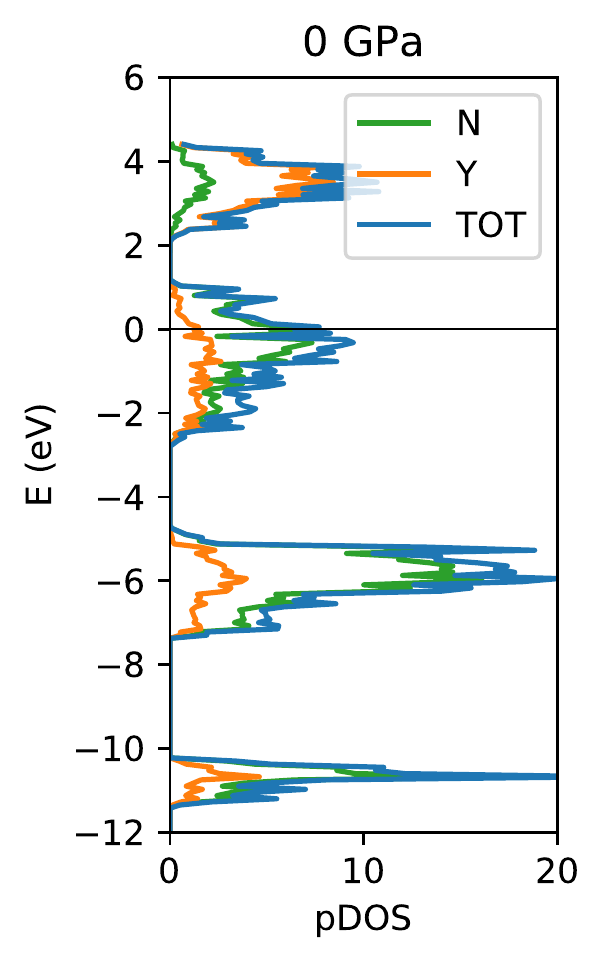} &
    \includegraphics[width=0.2\columnwidth]{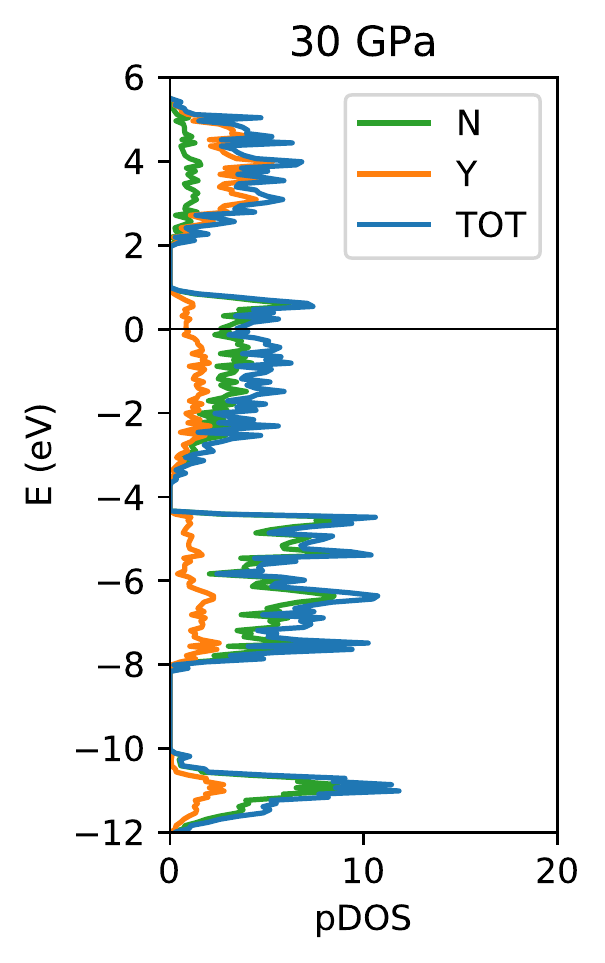} &
    \includegraphics[width=0.2\columnwidth]{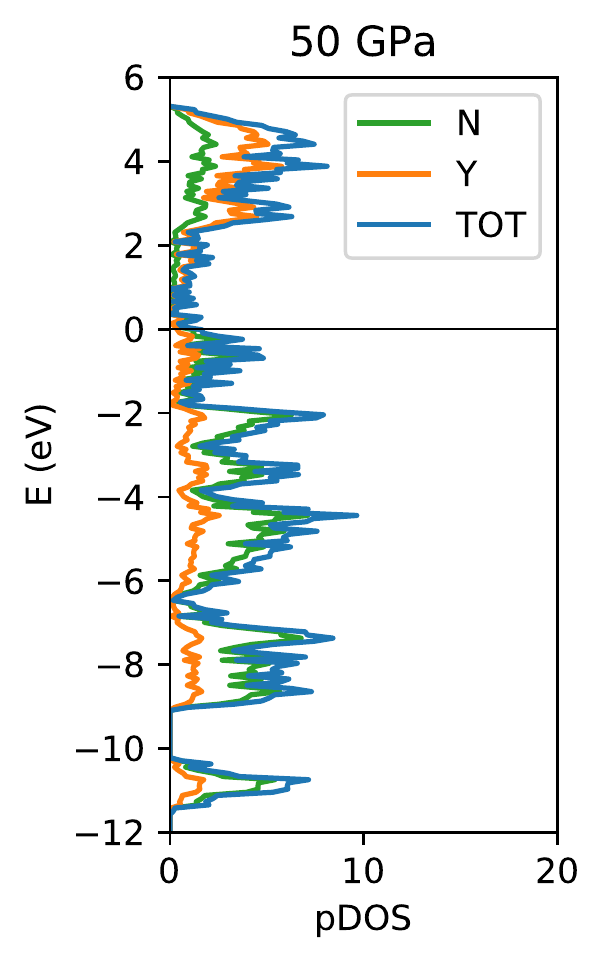} &
    \includegraphics[width=0.2\columnwidth]{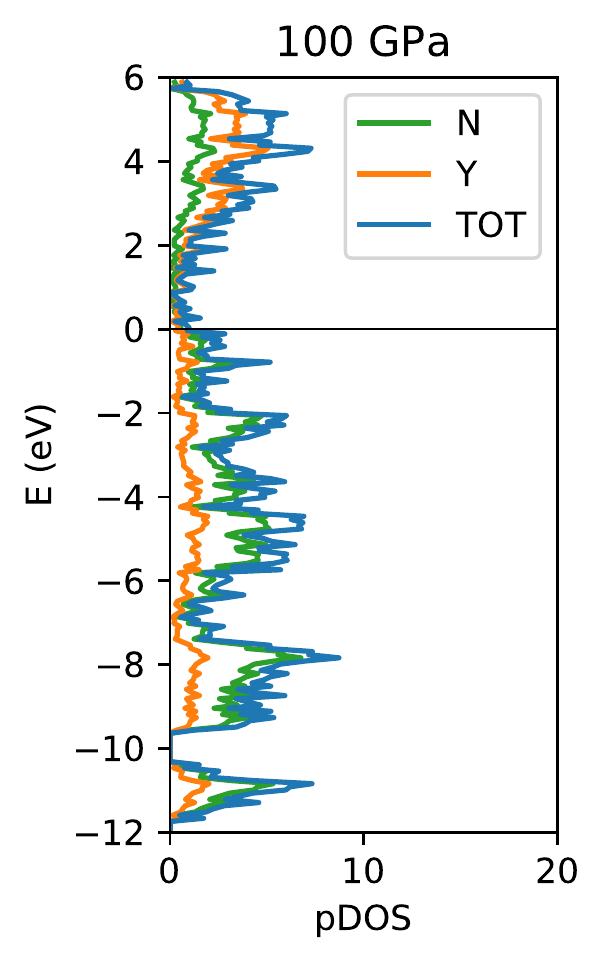} \\
    \end{tabular}
    \caption{Projected electronic density of states of selected polymorphs and pressure.
    (a) C2/c at 0~GPa. (b) P2$_1$/c\#1 at 30~GPa. (c) and (d) P2$_1$/c\#2 at
    50 and 100~GPa. The Fermi level is denoted by the black horizontal line.}
    \label{fig:DOS}
\end{figure}

In Fig.~\ref{fig:DOS} we report the projected electronic density of states of
selected structures, with respect to the Fermi level. The C2/c (0 GPa) and P2$_1$/c\#1
(30 GPa) structures are metallic and their valence band has a large N~$2p$ character.
The Y states instead contribute mainly to the empty states. From the electronic
structure point of view the formal valence of these polymorphs is closer to
Y$^{+2}$(N$_2)^{2-}$ rather than to Y$^{+4}$(N$_2)^{4-}$, i.e. closer to the
\emph{dinitride} structure. The presence of a unique N..N bond length indicates that
bond disproportionation (i.e. (Y$^{3+}$)$_2$(N$_2)^{2-}$(N$_2)^{4-}$) does not takes place. This
is conceivable since the electronic screening of the metallic state tends to
delocalize the charge carriers and suppresses the ionic character.

\begin{figure}
    \centering
    \includegraphics[width=0.3\columnwidth]{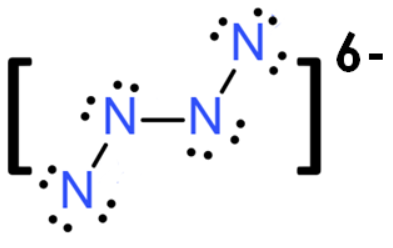}
    \caption{The simplest Lewis structure with only $\sigma$ bonds of the (N$_4$)$^{6-}$ moiety
    found in the P$2_1/c$\#2 structure. In reality the chemical bonding has a partial
    multicenter $\pi$ contribution.}
    \label{fig:lewis}
\end{figure}

The situation changes upon increasing the pressure above 50~GPa, following the formation of the
N$_4$ moieties. The electronic density of states is semimetallic, with a low density of
states up to $\sim$2~eV above the Fermi level. This can be explained by the formal
valence (Y$^{3+}$)$_2$(N$_{4}$)$^{6-}$, in which the N$_{4}$ moiety has 6 extra electrons
(i.e. 26 valence electrons). The 26 valence electrons (13 electron pairs) could be naively
arranged as 3 single bonds and 10 lone pairs as shown by the Lewis structure with single
N$-$N bonds as shown in Fig.~\ref{fig:lewis}. Moreover, the (N$_{4}$)$^{6-}$ moiety
fulfills the $6n+2$ Wade rule in $n$-member linear chains. However the Lewis structure
does not reflect the actual charge distribution and chemical bonding. In fact the (N$_{4}$)$^{6-}$
anion is planar, indicating a partial multi-center $\pi$ bonding. 
In principle this structure would possess a finite band gap, but the increased overlap due
to the external pressure tends to close the gap. Notice also the increase of the band
width of the occupied states and the larger hybridization between Y and N states. Whereas
the low pressure phases can be described as intermetallic, the high pressure phase could
be described more as a salt-like, Zintl phase. This hypothesis could be confirmed or disproved
by advanced chemical bond analysis.~\cite{Wagner2016,Outeiral2018}
Interestingly, a (N$_{4}$)$^{6-}$ anion in the planar $cis$ conformation is found in the USPEX
search but its enthalpy is larger than that of the $trans$ conformation.

\section{Discussion}
It is interesting to study the behavior of N..N bond length as a function of pressure and
compare to other di-nitrides. The ThC$_{2}$-like structure at ambient pressure has a N..N
bond length of 1.33~\AA, which is intermediate between a single and a double bond (see
Tab.~\ref{tab:bonds}). Indeed, when the cation is divalent
(like in the dinitrides as BaN$_{2}$), N$_{2}$ is better described in the form (N=N)$^{2-}$,
whereas if the cation valence is higher (like in the pernitrides as PtN$_{2}$), the N$_{2}$ moiety
is best described as (N$-$N)$^{4-}$. In this particular case, however, the situation is
more challenging to rationalize, since there is no evidence of the yttrium cation being either
divalent or tetravalent: it is always reported as trivalent. This consideration brings us to consider
two different options to rationalize the valence of these compounds: the first possibility is to
have static fluctuations in the N$_2$ bond lengths, where some N..N longer bonds are longer than
others averaging out to around 1.33~\AA; the second possibility is to have an electron forming a
partially filled band. The analysis of the DOS in Fig.~\ref{fig:DOS} seems to suggest the
second scenario. This has been already proposed to be the case of LaN$_{2}$~\cite{Wessel2010}.
After the first structural transition, the situation remains similar, with a N$-$N bond
length of 1.32~\AA.

\begin{table}
    \centering
    \begin{tabular}{|c|c|c|}
       \hline
         Structure & N$-$N bond length / \AA  & Ref.\\
         \hline
         BaN$_{2}$ & 1.23 & Ref.~\cite{Wessel2010} \\
         PtN$_{2}$ & 1.41 & Ref.~\cite{Wessel2010} \\
         OsN$_{2}$ & 1.43 & Ref.~\cite{Wang2011} \\
         IrN$_{2}$ & 1.30 & Ref.~\cite{Montoya2007} \\
         LaN$_{2}$ & 1.30 & Ref.~\cite{Wessel2010} \\
         YN$_{2}$, C2/c, 0~GPa & 1.33 & this work \\
         YN$_{2}$, P2$_1$/c\#1, 30~GPa & 1.32 & present work \\
         YN$_{2}$, P2$_1$/c\#2, 50~GPa & 1.42, 1.44 & present work \\
         YN$_{2}$, P2$_1$/c\#2, 50~GPa & 1.39, 1.40 & present work \\
         \hline
    \end{tabular}
    \caption{Nitrogen-nitrogen bond lengths in various pernitrides and dinitrides, reported from literature as well
    as calculated in the present work for YN$_2$, as a function of pressure.}
    \label{tab:bonds}
\end{table}
The situation changes dramatically after the second transition, at 40 GPa, with the
formation of N$_{4}$ chains, the bond lengths increase significantly with respect to lower
pressures: the central bond is 1.42~\AA\ and the two others are 1.44~\AA. Indeed, the (N$_{4}$)$^{6-}$
moiety can be described by a single Lewis formula with all single N$-$N bonds. As a consequence,
the system is a closed shell and the solid is semimetallic. At the very high pressure of 100
GPa the N-N bond lengths in N$_{4}$ moiety become 1.39 and 1.40 \AA.

Wessel and collaborators~\cite{Wessel2010} observed that mechanical hardness and bulk modulus
are larger in pernitrides than in dinitrides, due to the different character and filling of
N$_2$ orbitals.
We calculated the bulk modulus of the three YN$_2$ polymorphs by fitting to the Birch-Murnaghan
3rd order equation of state~\cite{Birch1947}. The results are reported in Tab.~\ref{tab:bulk}, and compared to
other dinitrides and pernitrides from literature. Indeed, the low pressure YN$_2$ phases display a
bulk modulus larger than the dinitride BaN$_2$ and similar to that of LaN$_2$. Interestingly,
upon the formation of N$_4$ chains, the bulk modulus increases by $\sim$60~GPa.

\begin{table}
    \centering
    \begin{tabular}{|c|c|c|}
       \hline
        Structure & Bulk Modulus (GPa) & Ref.  \\
         \hline
         BaN$_{2}$ & 46  & Ref.~\cite{Weihrich2003}\\
         SrN$_{2}$ & 65  & Ref.~\cite{Weihrich2003}\\
         PtN$_{2}$ & 256 & Ref.~\cite{Weihrich2003}\\
         OsN$_{2}$ & 358 & Ref.~\cite{Montoya2007} \\
         IrN$_{2}$ & 428 & Ref.~\cite{Wu2007} \\
         LaN$_{2}$ & 86  & Ref.~\cite{Weihrich2003}\\         
         YN$_{2}$, C2/c & 117 & present work \\
         YN$_{2}$, P2$_{1}$/c\#1 & 118 & present work \\
         YN$_{2}$, P2$_{1}$/c\#2 & 174 & present work \\
         \hline
    \end{tabular}
    \caption{Bulk moduli in various pernitrides and dinitrides, reported from literature as well
    as calculated in the present work for YN$_2$ with EosFit~\cite{eosfit} 
    (Birch-Murnaghan, 3rd Order).}
    \label{tab:bulk}
\end{table}

Regarding superconductivity one could estimate estimate $T_c$ from the BCS formula
$T_c = 1.134\, T_D\, \exp\left[1/(N_0\ g)\right]$ which links the superconducting temperature
to the Debye temperature $T_D$ (which is high in hard materials), to the density of states at
the Fermi level $N_0$ and to the average electron phonon coupling $g$.
Our dinitrides are not super-hard materials compared to transition metal 
mono-nitrides~\cite{Shy1973,Wang2015}.
There is no particular reason to expect a large electron phonon coupling and the density
of states at the Fermi level is not very large. As a consequence, we expect that YN$_2$ T$_c$
to be smaller than that of the mono-nitrides and much smaller that the superconducting
hydrides.~\cite{Semenok2020}

Our methods can be applied to study the formation and stability of the series of LnN$_2$
compounds (Ln=La..Lu). We expect that as the size of the cation is reduced the phase
transition will occur at higher pressure and new compact structures might be found. Lanthanide
cations with multiple oxidation states (i.e. Ce, Pr, Sm, Eu, Tb, Tm, Yb) will represent a
challenge for DFT and one must employ advanced techniques like dynamical mean field theory
(DMFT).~\cite{Georges1996}

\section{Conclusions}
We found by crystal structure prediction the most stable structures of YN$_2$ up to 100~GPa.
At low pressure, the system adopts the ThC$_2$ structure like LaN$_2$. Unfortunately this
structure is thermodynamically unstable with respect to the decomposition into YN and solid N$_2$
and dynamically unstable at ambient pressure.
This structure is characterized by N$_2$ dimers and, upon increasing pressure, it transforms
into a monoclinic polymorph (P2$_1$/1\#1) with no evident N$_2$ layers. By comparing the nitrogen
bond length and the bulk modulus, we arrive at the conclusion that both C2/c and P2$_1$/1\#1
predicted polymorphs can be categorized as \emph{dinitrides} and the N$_2$ chemical bond is
closer to a double N=N bond. Finally, above 40 GPa, N$_4$ moieties are formed and the structure
is semimetallic. We verified the dynamical stability of the two high pressure structures by
computing the phonon density of states. Our work is the first step towards studying the
formation and stability of LnN$_2$ compounds (Ln=La..Lu). In principle we can substitute
Y with any lanthanide in every structure generated my USPEX and compute the enthalpies
in the whole pressure range. This will provide a tentative phase diagram for each LnN$_2$
compound that will provide useful indication for future high pressure experiments.

\begin{acknowledgement}
We acknowledge financial support from the Center for Materials Crystallgraphy (CMC). We thank
Martin Bremholm and Carlo Gatti for helpful discussions. Calculations were performed at the
CINECA supercomputing center thanks to the ISCRA HP10C7BPGD, HP10C5WGQ7 and HP10CPXESJ grants.
\end{acknowledgement}

%

\bibliography{uspex_aflow,nitrides,DFT}  

\end{document}